\documentclass[doublecol]{epl2} 

\usepackage{graphicx}
\usepackage{float}
\usepackage{amsmath, amssymb}

\title{Importance of initial conditions in the polarization of complex networks}

\author{Snehal M. Shekatkar \and Sukratu Barve}
\shortauthor{S. Shekatkar\etal}

\institute{                    
  Centre for modeling and simulation, SP Pune University, Pune 411007
}

\abstract{
Currently used models of opinion formation use random initial conditions. In reality, most people in a social network, except for a small fraction of the population, are initially either unaware of, or indifferent to, the disputed issue. To explore the consequences of such specific initial conditions, we study the polarization of social networks when conflicting ideas arise on two different seed nodes and then spread according to a majority rule. Using the configuration model and the stochastic block model as examples, we show that this framework leads to substantially different outcomes than those which employ random initial conditions. Moreover, the empirically observed splits in the karate and the dolphins' networks naturally come out of this paradigm. Our work thus suggests that the existing opinion dynamics models should be reevaluated to incorporate the initial condition dependence. 
}

\begin{document}

\maketitle

\section{Introduction}\label{introduction}
When faced with a question with two conflicting answers, such as which candidate to vote for, or whether the real-world networks are scale-free \cite{broido2018scale, barabasi2018love}, social networks often polarize by forming two opinion groups. This emergence is explained by the binary opinion models and their generalizations as a result of the `majority rule' whereby the choice made by the majority of social acquaintances of a node, dictates the selection of its future choice.\cite{castellano2009statistical, shao2009dynamic, biswas2009model, gleeson2013binary, dandekar2013biased, hindes2017large, medina2017consensus, galam2015two, amato2017opinion}.

Despite being successful at providing many fascinating insights into the dynamics of social systems, these models assume that initially, every node is in one of the two opposite states. This assumption, however, is pretty unrealistic since in most cases, a sizable fraction of the population is initially either unaware of or indifferent to the disputed issue, and only a small number of people have a definite stand over it. Hence, it is essential to study how the `network polarizability' is affected by such non-random initial conditions. 

In this paper, we study a particular variation of this idea in which two chosen nodes, called seed nodes, are initially infected with opposite opinions while the remaining nodes are kept in a neutral state. The seed nodes then spread their opinions in the network such that every node changes its choice at each time step following a majority rule until a steady state is reached. Depending upon the selection of seed nodes, we either observe the complete consensus state or a highly polarized state, or a state with intermediate polarization. A similar model has been studied in \cite{zhao2014competitive} to explore the competition between two fixed nodes which never change their opinions. On the contrary, in our model, the seed nodes are not forced to retain their original opinions and are subjected to the same majority rule. Furthermore, since in real social networks one seldom knows the seed nodes on which the conflicting ideas are formed, we must be agnostic about the choice of the seed pair while talking about the polarizability of a given network. We thus propose to average the polarization over a large number of seed pairs to estimate the average polarizability of the network. The initial condition dependence in the opinion dynamics has also been investigated for a few particular situations such as for studying the effectiveness of interventions to change the adolescent smoking behavior \cite{adams2016initial}, and even in the case of `bounded-confidence' opinion models in the agent-based settings \cite{carro2013role}. Interestingly, the effect of inital conditions is well studied in the case of evolutionary game theory in which a ``prepared'' initial spatial distribution of strategies (or the distribution in the underlying metric space for networks) has been shown to lead to different results than those obtained using random initial distribution \cite{perc2017statistical, kleineberg2017metric, amato2017interplay}. However, the `seed initial conditions' (SICs) that we introduce here have not been considered in the literature as per our knowledge, and as we demonstrate in the ensuing sections, are central to the understanding of observed polarizations in social systems.

It is important to note here that though the three-state opinion models, similar to our model, are well studied in the literature in the context of polarization, they still use RICs and hence our work fundamentally differs from them \cite{mobilia2011fixation, crokidakis2013role, balenzuela2015undecided}. 

Along with the initial condition dependence, we also want to investigate how various structural features of a network contribute to the overall polarizability. Here, we focus on two most commonly found structural traits in social networks: a fat-tailed degree distribution and community structure \cite{fortunato2016community}. Since the asymptotic steady state in our model crucially depends on the choice of the seed nodes, any answer to such question must be given in terms of an average taken over a large number of seed pairs. As we show in the following, communities tend to make the network more polarized as expected, while the existence of high degree nodes directs it towards consensus states. All the numerical simulations are carried out using graph-tool \cite{peixotograph-tool2014}.

\section{\label{simple_opinion}A simple model of opinion formation}
Consider a very general model of opinion formation as given below.

\begin{equation}
\label{opinion_general}
    {\bf x}(t+1) = {\bf f}({\bf x}(t), {\bf A}(t), \{D\})
\end{equation}

Here ${\bf x} = (x_1,x_2,\cdots,x_n)^T$ is the vector representing the opinion values on the nodes and $f$ is a function that connects the states at times $t$ and $t+1$ and can have arbitrary form. Also, ${\bf A}$ is the adjacency matrix of the network that in general is asymmetric with any real numbers (including negative) as its entries to represent the strengths of the connections between the nodes. Moreover, these entries can, in general, be functions of time. The set $\{D\}$ collectively represents the remaining parameters of the model.

However, our aim in this paper is not to model the opinion formation process with as much realism as possible. Rather, we want to see how the polarization dynamics is affected when the initial conditions are in the form of two seed nodes with opposite opinions instead of being completely random. To achieve this, we consider a simple version of the model given by eq.(\ref{opinion_general}). The two opposite opinions can be conveniently modeled as $+1$ and $-1$ while the neutral view can be represented by $0$. Also, in this paper we focus on undirected networks and update the states according to the ``majority rule'' as follows:

\begin{equation}
\label{opinion_simple}
        x_i(t+1) =  \text{sgn}\left(x_i(t) + \sum\limits_j A_{ij}x_j(t)\right)
\end{equation}
where sgn is the sign function that takes value $+1$ when its argument is positive, $-1$ when its argument is negative and $0$ otherwise. The states of all the nodes are updated simultaneously. We mention that a similar model has been given in \cite{shao2009dynamic}, but in that case, the initial state of the network was taken to be a random state. This model is also reminiscent of the label propagation method for the detection of communities in networks \cite{raghavan2007near} except that apart from the states of the neighboring nodes, the present state of the node is also taken into account.  

\subsection{Polarization index}
The dynamics of eq.(\ref{opinion_simple}), after a few iterations, results in states which do not change further. Hence we restrict ourselves to only such steady states. The network is highly polarized if the numbers of nodes in at least two of the three states are roughly equal. Nonetheless, for all practical purposes, we can talk in terms of $-1$ and $+1$ states since the steady states with a large number of $0$ node values are rare. Thus, to quantify the polarization, we define the following index:

\begin{equation}
r = 1 - 4(n^--0.5)^2
\end{equation}

Here, $n^-$ is the fraction of nodes with negative states. It can be easily verified that for the consensus or unpolarized states, for which $n^- \approx 0$ or $n^- \approx 1$, the polarization index $r \approx 1$ whereas for the highly polarized network states, $r \approx 1$.

\section{Effect of initial conditions}\label{effect_init}
\subsection{Power-law configuration model}\label{config}
\begin{figure}
        \includegraphics[width=0.45\textwidth]{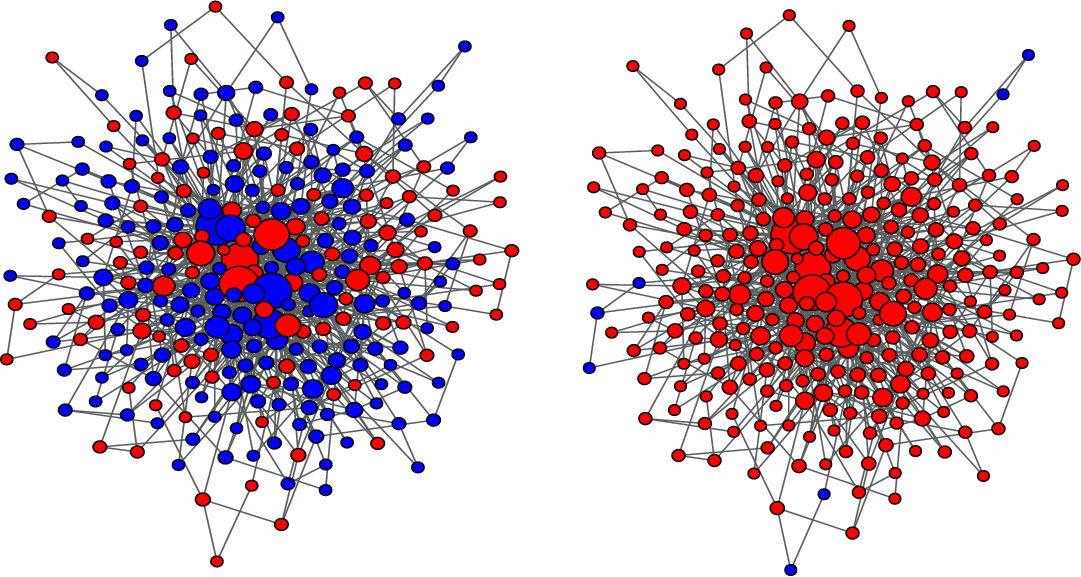}
        \caption{\label{config_states} Different seed pairs lead to different steady states. The presented network is drawn from the configuration model with a power-law degree distribution $k^{-\alpha}$}
\end{figure}

We start by studying the effect of a fat-tailed degree distribution on the network polarization when there are two seed nodes. Thus, as a representative model of such networks, we consider the configuration model \cite{newman2010networks} with a degree sequence drawn from the power-law distribution $p(k) \sim k^{-\alpha}$ and explore the results of running the dynamics of eq.(\ref{opinion_simple}) on it. In the configuration model, the degree value for each node of the network is specified by assigning a certain number of half-edges to it, and these half-edges are then randomly connected to each other. fig.~\ref{config_states} shows two different states obtained for the power-law configuration model. As we can see, different choices of the seed nodes can lead to drastically different steady states. 

\begin{figure}
        \includegraphics[width=0.45\textwidth]{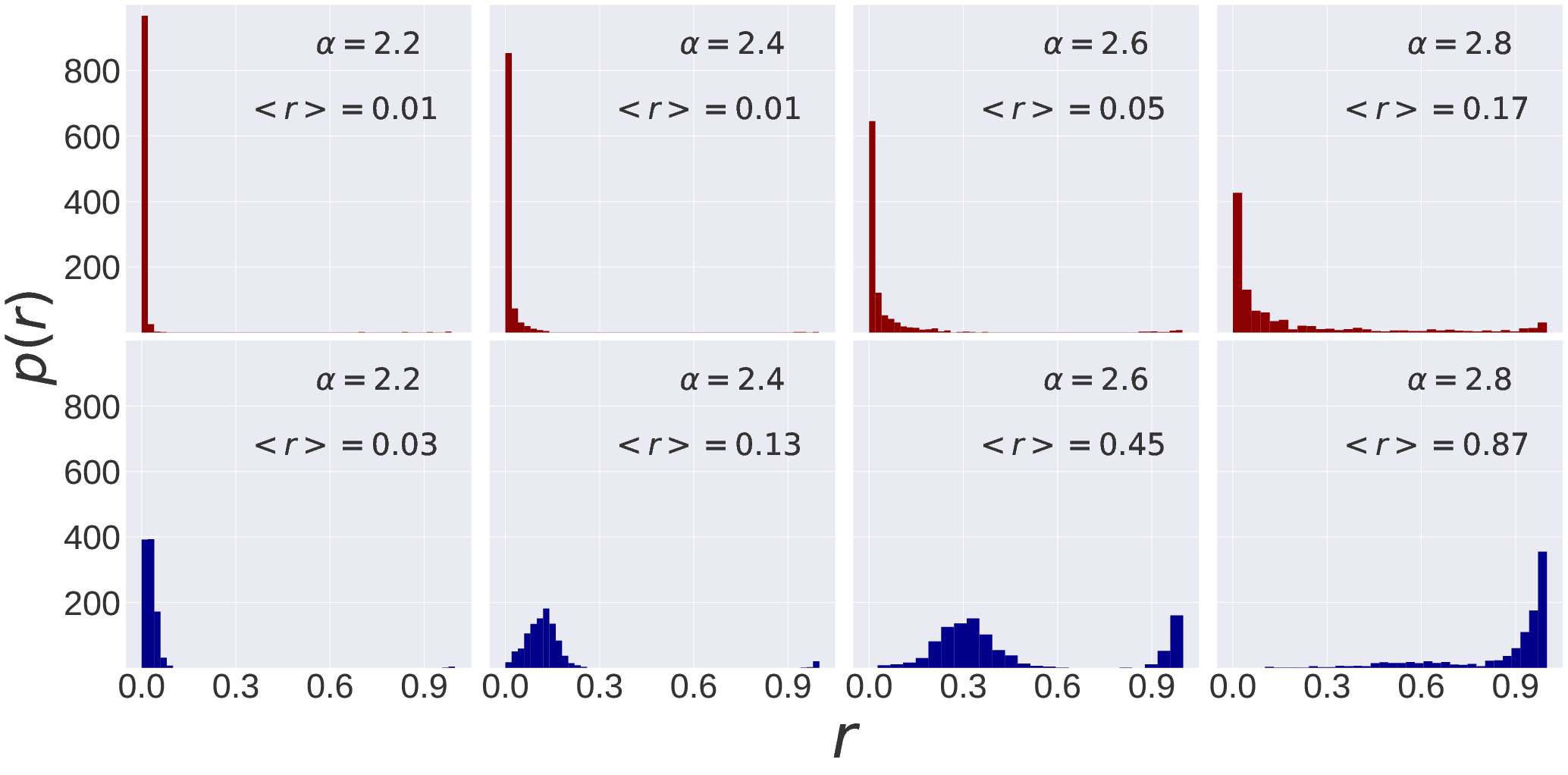}
        \caption{\label{config_pol_hists} Histograms of the polarization $r$ for different scaling indices $\alpha$ for the configuration model of size $N=5000$ with $1000$ realizations for each. Top (brown): the initial conditions are chosen in the form of seed nodes (SIC). Bottom (blue): the initial conditions are random (RIC). The difference in the polarization distributions for the two types of initial conditions is evident. The number of bins is calculated using Freedman-Diaconis rule \cite{freedman1981histogram}.}
\end{figure}
How does the abundance of hubs affect the emergence of high polarization states? In the power-law configuration model, this abundance can be tuned by varying the scaling index $\alpha$. In fig.~\ref{config_pol_hists}, we show histograms of polarization values $r$ for different power-law scaling indices $\alpha$ with seed initial conditions (SICs henceforth). In the same figure, we show the histograms obtained starting with the random initial conditions (RICs henceforth) so that each node is in one of the two states $+1$ or $-1$ with an equal probability. In both cases, since as $\alpha$ is increased, the number of high degree nodes decreases, an emergence of high-polarization states becomes more probable. This is understandable since if the network contains several high degree nodes, they dominate the network with their opinion reducing the chances of observing polarized states. However, for a given value of $\alpha$, the corresponding histograms in two cases can be seen to be quite different from each other. In particular, the average value $<r>$ of the polarization can be seen to be consistently higher with RIC.

\subsection{Stochastic block model (SBM)}\label{sbm}

Apart from heavy-tailed degree distributions, another common structural feature seen in almost all the social networks is an existence of community structure \cite{girvan2002community, karrer2011stochastic,lusseau2003bottlenose,peixoto2014hierarchical}. A community in a complex network is defined as a group of nodes that have the same connection probabilities to the other nodes in the network. In particular, social networks exhibit an assortative type of communities so that the connection probabilities inside the groups are higher than the probabilities between the groups. We want to see how the existence of assortative communities affects the spread of conflicting opinions that are originated on two different nodes. 

A straightforward way to study this question is to construct graphs with planted or ``hand-made'' communities and then simulate the dynamics on them. There exist several different random graph models which contain the idea of communities or blocks in their description \cite{rosvall2008maps,karrer2011stochastic,peixoto2012entropy,peixoto2013parsimonious}. Arguably, the simplest of these is the famous `Stochastic block model' or SBM for brevity \cite{peixoto2012entropy}. In the traditional SBM, we start with $N$ nodes and group them into a $B$ number of blocks or modules. Then every pair $(i,j)$ of nodes is connected with a probability $\omega_{b_ib_j}$ where $b_i$ and $b_j$ denote the blocks to which the nodes $i$ and $j$ belong to respectively. For generating a highly assortative community structure that we are interested in, we make the probabilities $\omega_{rr}$ inside the groups significantly higher than the probabilities $\omega_{rs}$ between the groups.

The planted partition model is a special case of the stochastic block model described above. To construct it, we set all the inter-block probabilities $\omega_{rs}$ to the same value $\omega_{\text{out}}$, and all the probabilities inside the groups $\omega_{rr}$ to the same value $\omega_{\text{in}}$. For generating a strongly assortative community structure that we are interested in, we set $\omega_{\text{in}} >> \omega_{\text{out}}$. 

Consider the SBM with a Poisson degree-distribution. In such network, the degree values of all the nodes are concentrated around the average value. Thus, unlike the power-law configuration model, there are no `hubs' which can dominate the network. The model is therefore ideal if we want to study effects of community structure only. 

\begin{figure}
        \includegraphics[width=0.45\textwidth]{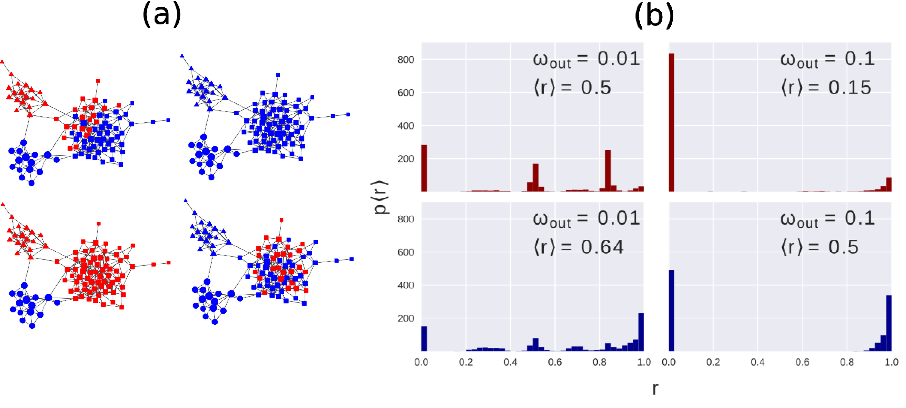}
        \caption{\label{four_states_3blocks}(a) Different steady states for the Poisson planted partition model. (b) Polarization histograms for SBM with SICs (top panel, brown) and with RICs (bottom panel, blue) illustrate that the two types of initial conditions lead to markedly different outcomes. See text for details.}
\end{figure}

In fig.~\ref{four_states_3blocks} (left) we show some of the steady states for the Poisson SBM with one large and two small blocks. Interestingly, the nodes in the same community, despite being densely connected to each other, can have opposite states asymptotically as the picture shows. In the same figure, we also show the polarization histograms with and without random initial conditions obtained using $1000$ random realizations for a much larger network ($N=5000$) with $3$ communities. The relative community sizes are $(0.7, 0.15, 0.15)$ and distributions for $\omega_{\text{out}} = 0.01$ and $\omega_{\text{out}} = 0.1$ are shown with $\omega_{\text{in}} = 0.7$ fixed. The average degree $c = 6$ in all the cases. Similar to the case of configuration model, the random initial conditions give different results. 

\begin{figure}
    \includegraphics[width=\columnwidth, trim = 30 50 0 0, clip = true]{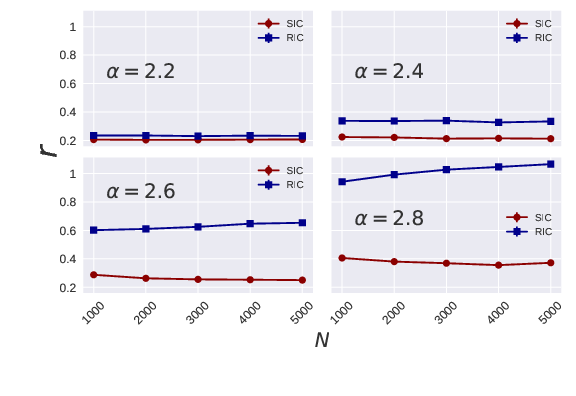}
    \caption{\label{pol_vs_size} Variation of the polarization $r$ with the size $N$ of the network for different $\alpha$ values for the power-law configuration model. SICs (red circles) are seen to produce significantly lower average polarization than RICs (blue squares). Values are averaged over $1000$ random realizations.}
\end{figure}

\begin{figure}
    \includegraphics[width=\columnwidth, trim = 30 0 0 0, clip = true]{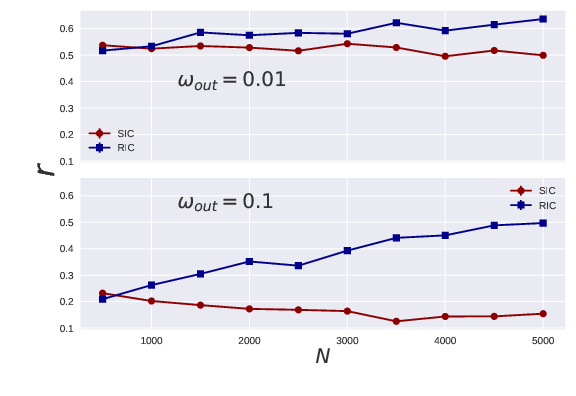}
    \caption{\label{pol_vs_size_sbm} Variation of the polarization $r$ with the size $N$ of the network for different $\omega_{\text{out}}$ for the poisson SBM of fig.~\ref{four_states_3blocks} averaged over $1000$ realizations. Akin to the configuration model, SICs produce lower average polarization.}
\end{figure}

We now wish to see how the polarization is affected by the size of the network. We start with the power-law configuration model. In fig.~\ref{pol_vs_size}, we show variations of $r$ with the size $N$ of the network. As we can see, variations are starkly different for RICs and SICs, especially for larger $\alpha$ values for which random initial conditions predict a substantial increase in the average polarization whereas the seed initial conditions predict the exact opposite. For the Poisson SBM, we see qualitatively similar results as shown in fig.~\ref{pol_vs_size_sbm}.

\section{Degree-corrected SBM}\label{karrer-newman}
The configuration model and the stochastic block model encode two different topological features of complex networks. The former controls only the degree distribution while the later allows tuning the nature of community structure. Thus, to be more realistic, we need a model in which both the aspects could be accommodated. In particular, we want to capture a crucial aspect of many real-world networks, that of the degree heterogeneity, along with the presence of communities. In several real-networks, the degrees of the nodes are seen to be taking values in a vast range. In other words, the degree distributions of these networks are not peaked around their average values. On the contrary, the simple SBM produces graphs with the Poisson degree distribution so that most of the nodes have degrees around the average of the distribution. Karrer and Newman have extended the traditional stochastic block model to incorporate the arbitrary degree distributions \cite{karrer2011stochastic}, and it is known as the `degree-corrected SBM'. In their model, along with the connection probabilities between the groups, each node $i$ is endowed with a parameter $\theta_i$ that is proportional to its specified degree. The degree values can be drawn from any distribution and then the pair $(i,j)$ is connected with the probability proportional to $\theta_i\theta_j\omega_{b_ib_j}$. 

In the most general setting, one can vary all the elements $\omega_{rs}$ of the $B\times B$ probability matrix and see how the resultant network states change. However, since our focus here is only on understanding the effect of communities on the asymptotic states, we simplify the situation by setting $\omega_{rr} = \omega_{\text{in}}$ and $\omega_{rs} = \omega_{\text{out}}$ when $r \neq s$. This is known as the degree-corrected planted partition model \cite{newman2016equivalence}. As mentioned earlier, many real-networks possess fat-tailed degree distributions. As a consequence of this, in these networks, there exist few nodes with exceedingly large degree-values compared to the average degree of the network. One such distribution is the power-law distribution $p(k) \sim k^{-\alpha}$ where $\alpha$ is the scaling-index, and we will use it as a ``proxy'' for representing the fat-tailed distributions. 

\begin{figure}
    \includegraphics[width=0.95\columnwidth]{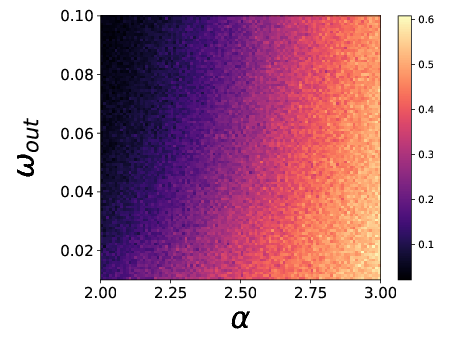}
    \caption{\label{alpha_pout_plane}Heatmap of average polarization in $\alpha$-$\omega_{\text{out}}$ plane with $\omega_{\text{in}} = 0.7$ for the Karrer-Newman model. For each $\alpha$-$\omega_{\text{out}}$ combination, polarization values of $200$ realizations are averaged for networks of size $N = 100$.}
\end{figure}

The results of running the dynamics on networks generated using the degree-corrected SBM with the power-law degree distribution and three blocks with a fraction of nodes in each block equal to $(0.7, 0.15, 0.15)$ are summarized in the form of heatmap of polarization in fig.~\ref{alpha_pout_plane}. As the figure shows, increasing $\alpha$ (which amounts to the decrease in the abundance of hubs) increases the polarization while increasing $\omega_{\text{out}}$ (which amounts to weakening the community structure) decreases the polarization. In other words, degree-distribution and community structure act as controlling features for polarization. The region in the upper left corner represents a region with a close to zero polarization and corresponds to a high abundance of hubs and very weak community structure. On the other hand, a high polarization exists in the lower right corner because of a lower abundance of hubs and strong community structure. All the results in this plot correspond to SICs.


\section{Analysis of the stability of the final state}\label{stability}
So far we did not consider the question of stability of the asymptotic opinion communities. We want to investigate the effect of a small perturbation to the asymptotic state of the network. This can be done using linear stability analysis. The $sgn(x)$ function in eq.(\ref{opinion_simple}) is discontinuous at $x=0$ and hence is difficult to handle for this purpose. Thus, we replace it with a smooth function which essentially captures the idea of jump at $x = 0$. With this modification, our model is given in eq.(\ref{opinion_arctan}).

\begin{equation}
\label{opinion_arctan}
x(n+1) = \frac{2}{\pi}\tan^{-1}\left[B\left(x_i(t) + \sum\limits_j A_{ij}x_j(t)\right)\right]
\end{equation}
As can be easily seen, this model is equivalent to the original model in eq.(\ref{opinion_simple}) in the limit $B\rightarrow \infty$. We have also numerically verified that the two models agree almost perfectly with $B=100$.

\begin{equation}
\label{Jacobian_ij}
    \begin{aligned}
        \left[\frac{\partial f_i}{\partial x_j}\right]_{x^{\ast}} &= \frac{2B}{\pi}\frac{A_{ij} + \delta_{ij}}{1 + B^2\left(x_i^\ast + \sum\limits_jA_{ij}x^{\ast}_{j}\right)^{2}}\\
&= \frac{2B}{\pi}\frac{A_{ij} + \delta_{ij}}{1 + \tan^{2}(\frac{\pi}{2}x^{\ast}_i)}\\
&= \frac{2B}{\pi}\cos^2\left(\frac{\pi}{2}x_i^{\ast}\right)\left(A_{ij} + \delta_{ij}\right)
    \end{aligned}
\end{equation}
This can be written as a matrix equation:

\begin{equation}
\label{J_matrix}
{\mathbf J} = {\mathbf D}({\mathbf A}+{\mathbf I})
\end{equation}
where ${\mathbf D}$ is diagonal matrix with: 

\begin{equation}
D_{ii} = \frac{2B}{\pi}\cos^2\left(\frac{\pi}{2} x_{i}^{\ast}\right)
\end{equation}
A given state ${\mathbf x}^{\ast}$ will be stable if all the eigenvalues of ${\mathbf J}$ have absolute values less than $1$. 

Using eq.(\ref{opinion_arctan}), we write the fixed point equation as follows:

\begin{equation}
x_i^\ast + \sum\limits_j A_{ij}x_j^\ast = \frac{1}{B}\tan\left(\frac{\pi}{2}x_i^\ast\right)
\end{equation}
Consider the asymptotic states for which $x_i=1$ or $x_i=-1$ only. For these states, L.H.S. is finite and non-zero and hence, when $B\rightarrow\infty$, $\tan\left(\pi x_i^\ast/2\right)\sim B$. Therefore, using eq.(\ref{Jacobian_ij}) and eq.(\ref{J_matrix}), one can see that the operator norm $||{\mathbf D}||\rightarrow 0$ as $B\rightarrow \infty$. The operator norm of a square matrix is the maximum among the absolute values of its eigenvalues, and so directly relates to the stability of a given state. Noting that the eigenvalues of ${\mathbf A}+{\mathbf I}$ do not change with $B$, we see that $||{\mathbf J}|| \leq ||{\mathbf D}||\cdot ||{\mathbf A}+{\mathbf I}||$ implies that $||{\mathbf J}|| \rightarrow 0$ as $B\rightarrow \infty$. Thus, the maximum value of the modulus of eigenvalues of ${\mathbf J}$ could be made smaller than $1$ by choosing a large enough $B$. This shows that all the asymptotic states for which $x_i=1$ or $x_i = -1$ are stable. 

On the other hand, if we consider the states having one or more zero values on the nodes, the above argument would not hold. In that case, we consider a general element of the Jacobian matrix eq.(\ref{Jacobian_ij}). Note that, $A_{ii} + \delta_{ii} = 1$ for all $i$. Selecting $i$ to be a node with a zero value, we obtain $\partial f_i/\partial x_i = (2B/\pi) \rightarrow \infty$  as $B\rightarrow \infty$. Thus, if we examine, the Jacobian matrix under the supremum norm, then we find that the norm of the Jacobian diverges as $B\rightarrow\infty$. For finite dimensional matrices, the supremum norm is equivalent to the operator norm (in fact to any other norm). So, the operator norm, and consequently, the eigenvalue with largest absolute value diverges with $B$. Thus, such states would be unstable. This is reasonable since a small perturbation would push the nodes with $x=0$ to either $+1$ or $-1$.

\section{Empirical networks}\label{examples}
\begin{figure}
\includegraphics[width=0.45\textwidth]{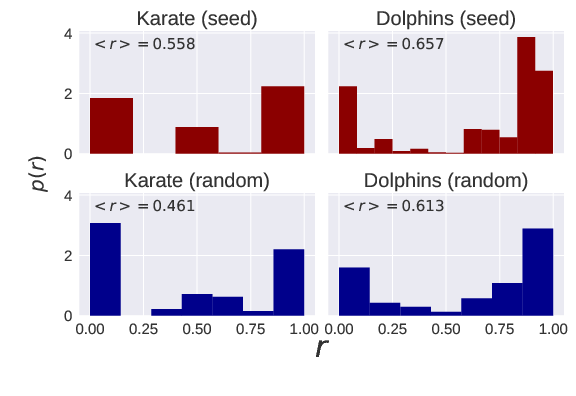}
\caption{\label{empirical} Top: Normalized histograms of polarization values obtained by using all possible seed pairs in the Zachary karate club network and the social network of bottlenose dolphins. Bottom: The corresponding histograms with RICs. In both the cases, RICs are seen to produce smaller average polarization than SICs. }
\end{figure}

We now apply the framework discussed so far to two well known empirical networks. The first one is the Zachary karate club network which is known to have got polarized and split into two parts. Underlying possible community structure is believed to be the major reason behind the split \cite{newman2006modularity, riolo2017efficient}. The ideas presented here suggest an alternative possibility that the observed polarization could be a repercussion of SICs since the point of the dispute was about the raising of the fees of the club, a problem with binary choice. Similar splitting has also happened in a social network of bottlenose dolphins from New Zealand \cite{lusseau2003bottlenose, newman2013spectral}. Though it is somewhat speculative to think about a dispute in case dolphins, we argue that since dolphins can make friendships, the possibility of a dispute arising in them should not be overruled. We apply our model with seed type initial conditions to these networks and obtain the average polarization by averaging over all possible seed pairs in these networks. 

fig.~\ref{empirical} shows the polarization distributions for these two networks with SICs and RICs. In this case, it is seen that the RICs give rise to lower average polarization than the corresponding SICs. Paradoxically, this is precisely opposite to the results of the previous sections in which RICs always produced higher polarization. We argue that this is a small size effect. In fact, we observe similar result for the Poisson SBM when the size of the network is small ($N \approx 100$). 

The present formalism, in fact, provides a much deeper insight into the social networks as we discuss now. The idea is to average over a large number of asymptotic states to find out whether the absolute difference between the values on the nodes directly connected by an edge, tends to be larger or smaller on an average. For an edge $e_{ij}$ that connects nodes $i$ and $j$, this average difference is:

\begin{equation}
    \triangle = \langle |x_i - x_j|\rangle
\end{equation}

\begin{figure}
\includegraphics[width=\columnwidth, trim = 0 30 0 50, clip = true]{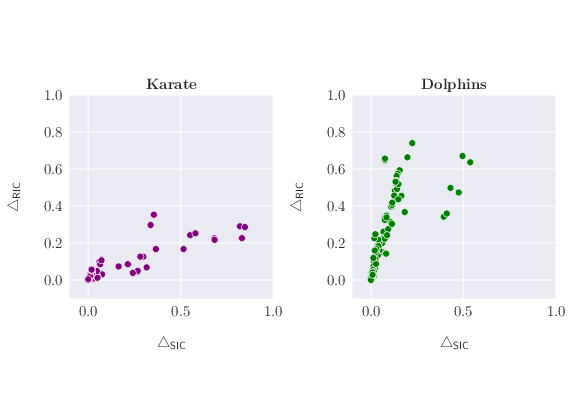}
\caption{\label{contrast_correlation}Scatter-plots of the RICs differences and the SICs differences for the Zachary karate network (left), and the social network of bottlenose dolphins (right). As the plots show, though the values are correlated, the individual points do not fall on the line $x=y$.}
\end{figure}

Here the angled brackets indicate an average over a large number of asymptotic states. We then ask whether the difference for a given edge obtained using SICs ($\triangle_{\text{SIC}}$) is approximately equal to the difference obtained using RICs ($\triangle_{\text{RIC}}$). At first thought, one may conclude that a SICs difference would be equal since the same network structure decides them. To check this, we make a scatter-plot of SICs difference and the RICs difference as shown in fig.~\ref{contrast_correlation} for the karate network and the dolphin network. Indeed, the two are highly correlated as expected. Nonetheless, it is clear from the plots that they do not tend to fall on the $x = y$ line. In other words, an edge for which RICs difference is small need not have small SICs difference, and the edges with a significantly high difference would, in general, be disparate depending upon whether SICs are used, or RICs are used. This result has significant implications for the polarization of networks as explained below. 

\begin{figure}
\includegraphics[width=0.9\columnwidth]{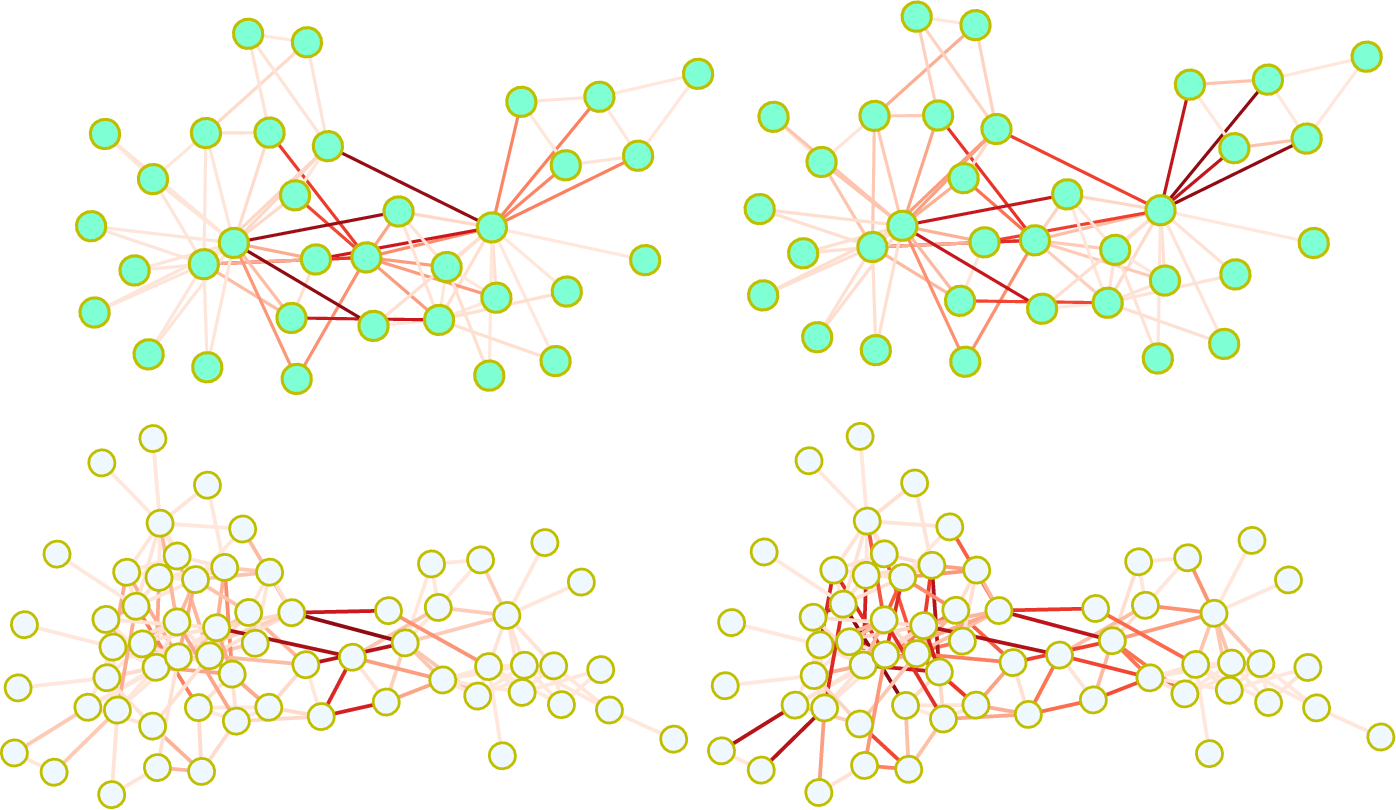}
\caption{\label{empirical_diff} Top: the Zachary karate network with the edges with higher difference represented by darker shades, obtained by averaging over SICs (left) and RICs (right). For SICs, darker edges are seen to lie in the ``middle'' whereas, for the RICs, they are scattered. Bottom: Similar picture shown for the dolphins' network for which SICs (left) predict darker edges perfectly in the middle, approximately the location at which the network was split.}
\end{figure}

In fig.~\ref{empirical_diff}, we show the karate network and the dolphins' network with the edges color-graded according to the average $x$-difference across them; darker edges have higher differences. On the left side of the figure, the differences are obtained using SICs whereas, on the right, they are obtained using RICs. Immediately, we see that SICs result into darker edges at the ``middle'' of these networks whereas darker edges are scattered when RICs are used. Most of these `high SICs difference edges' are the edges which broke during the observed splitting of these networks. The effect is particularly pronounced in the case of dolphins where removal of these edges, in fact, breaks the network completely into two parts, and the predicted breaking is almost similar to the observed breaking. This indicates that in both the networks, splitting could be a direct effect of SICs and that the initial conditions are as important as the network structure for predicting the polarization in social systems.

\section{Conclusions}
In this paper, we showed that the results of opinion dynamics models on networks could be unusually sensitive to the initial conditions. A possible reason behind this difference could be that the set of allowed initial conditions is now severely restricted and does not represent a truly random sample of the whole phase space. Moreover, we argued that for the polarization of complex networks, random initial conditions or RICs, in which each node is initially in one of the two opposite states with equal probability, are not realistic, and the seed initial conditions or SICs should be preferred. We also showed that SICs allow us to predict empirically observed polarization of networks like the karate club and the social network of dolphins with high accuracy. 

Some of the obvious generalizations of the work presented here include using more sophisticated models that incorporate stochasticity and using weighted and directed graphs. Also, it would be interesting to see how the results are affected by varying the community structure in several ways like the number of blocks, their sizes, and the overlaps. We anticipate that this initial condition dependence would be explored in more depth in the opinion dynamics studies in future to produce more realistic predictions. In particular, we expect that SICs formalism will be applied to more realistic models of opinion dynamics to see whether they produce results that have not yet been explained in the empirical social networks.  
\acknowledgments
S.M.S. thanks Dr. Mihir Arjunwadkar for many helpful discussions. S.M.S. acknowledges the funding from the National Post Doctoral Fellowship (NPDF) of DST-SERB, India, File No: PDF/2016/002672.


\end{document}